\documentclass[preprint,aps,superscriptaddress]{revtex4}

\usepackage{graphicx}
\usepackage{subfigure}
\usepackage{epsfig}
\usepackage{dcolumn}
\usepackage{bm}
\usepackage{ulem}
\usepackage{color}
\usepackage{textcomp}
\usepackage{indentfirst}
\usepackage{epstopdf}

\begin{document}

\title{Direct observation of local Rashba spin polarization and spin-layer locking in centrosymmetric monolayer PtSe$_2$}

\author{Wei Yao}
\affiliation{State Key Laboratory of Low Dimensional Quantum Physics and Department of Physics, Tsinghua University, Beijing 100084, China}

\author{Eryin Wang}
\affiliation{State Key Laboratory of Low Dimensional Quantum Physics and Department of Physics, Tsinghua University, Beijing 100084, China}

\author{Huaqing Huang}
\affiliation{State Key Laboratory of Low Dimensional Quantum Physics and Department of Physics, Tsinghua University, Beijing 100084, China}

\author{Ke Deng}
\affiliation{State Key Laboratory of Low Dimensional Quantum Physics and Department of Physics, Tsinghua University, Beijing 100084, China}

\author{Mingzhe Yan}
\affiliation{State Key Laboratory of Low Dimensional Quantum Physics and Department of Physics, Tsinghua University, Beijing 100084, China}

\author{Kenan Zhang}
\affiliation{State Key Laboratory of Low Dimensional Quantum Physics and Department of Physics, Tsinghua University, Beijing 100084, China}

\author{Taichi Okuda}
\affiliation{Hiroshima Synchrotron Radiation Center (HSRC), Hiroshima University, 2-313 Kagamiyama,
Higashi-Hiroshima 739-0046, Japan}

\author{Linfei Li}
\affiliation{Beijing National Laboratory for Condensed Matter Physics, Institute of Physics, Chinese Academy of Sciences, Beijing 100190}

\author{Yeliang Wang}
\affiliation{Beijing National Laboratory for Condensed Matter Physics, Institute of Physics, Chinese Academy of Sciences, Beijing 100190}
\affiliation{Collaborative Innovation Center of Quantum Matter, Beijing, P.R. China}

\author{Hongjun Gao}
\affiliation{Beijing National Laboratory for Condensed Matter Physics, Institute of Physics, Chinese Academy of Sciences, Beijing 100190}
\affiliation{Collaborative Innovation Center of Quantum Matter, Beijing, P.R. China}

\author{Chaoxing Liu}
\affiliation{Department of Physics, The Pennsylvania State University, University Park, Pennsylvania 16802-6300, U.S.A.}

\author{Wenhui Duan}
\affiliation{State Key Laboratory of Low Dimensional Quantum Physics and Department of Physics, Tsinghua University, Beijing 100084, China}
\affiliation{Collaborative Innovation Center of Quantum Matter, Beijing, P.R. China}

\author{Shuyun Zhou}
\altaffiliation{Email: syzhou@mail.tsinghua.edu.cn}
\affiliation{State Key Laboratory of Low Dimensional Quantum Physics and Department of Physics, Tsinghua University, Beijing 100084, China}
\affiliation{Collaborative Innovation Center of Quantum Matter, Beijing, P.R. China}

\date{Feb. 18, 2016}

\begin{abstract}

{\bf The generally accepted view that spin polarization is induced by the asymmetry of the global crystal space group has limited the search for spintronics \cite{Spintronics} materials to non-centrosymmetric materials. Recently it has been suggested that spin polarization originates fundamentally from local atomic site asymmetries \cite{ZungerNatPhys}, and therefore centrosymmetric materials may exhibit previously overlooked spin polarizations. Here by using spin- and angle-resolved photoemission spectroscopy (spin-ARPES), we report helical spin texture induced by local Rashba effect (R-2) in centrosymmetric monolayer PtSe$_2$ film. First-principles calculations and effective analytical model support the spin-layer locking picture: in contrast to the  spin splitting in conventional Rashba effect  (R-1), the opposite spin polarizations induced by R-2 are degenerate in energy while spatially separated in the top and bottom Se layers. These results not only enrich our understanding of spin polarization physics, but also may find applications in electrically tunable spintronics. 
}
\end{abstract}

\maketitle

\begin{figure*}
\includegraphics[width=13 cm] {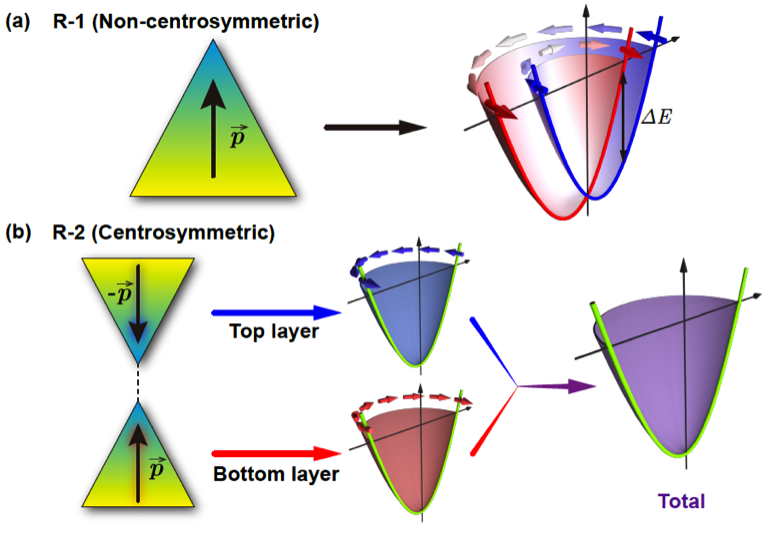}
\caption{\textbf{Illustration of conventional Rashba (R-1) and local Rashba (R-2) effects with different symmetries.} \textbf{a}, R-1 Rashba effect in non-centrosymmetric materials. The inversion symmetry is broken and there is a net dipole field. The right graph is the typical spin-resolved band structure induced by R-1 Rashba effect with a spin splitting in energy. \textbf{b}, R-2 Rashba effect in centrosymmetric materials. The inversion symmetry is preserved and there is a site dipole field in spite of the vanishing total dipole field. The right graphs illustrate the spin texture of R-2 Rashba effect. The two kinds of spins with opposite directions are spatially separated in different layers, resulting in overall zero spin polarization.}
\end{figure*}

The new insight that spin polarization in nonmagnetic materials originates from relativistic spin-orbit coupling (SOC) from the local asymmetry (atomic site group) \cite{ZungerNatPhys, SpinNewsViews, KingNatPhys, SpinNewsViews2} rather than the global asymmetry (bulk space group) has revolutionized our understanding of spin polarization physics. This has led to two forms of hidden spin polarization in centrosymmetric materials, R-2 (by site dipole field) and D-2 (by site inversion asymmetry) to be distinguished from the conventional Rashba (R-1, by dipole field)  \cite{Rashba, RashbaRev, LaShellAu, Bi111, Bi111K, Ir111, IshizakaBiTeI, BahramyPRB}  and Dresselhaus (D-1, by bulk inversion asymmetry) \cite{Dreselhauss} effects previously discovered in non-centrosymmetric materials. Different from the R-1 effect with a net spin polarization, the R-2 effect is characterized by compensated spin polarizations of opposite signs that are spatially segregated into two real space sectors forming the inversion partners, e.g. top and bottom layers (see schematic cartoon in Fig.~1). Compared to the R-1 effect with a large internal electric field which is difficult to be reversed by an external field, the spins induced by R-2 might have advantages  for electrically tunable spintronics devices due to the easy manipulation via the application of an external electric field  \cite{LaOBiS2NL, ZungerPRB}. Layered transition metal dichalcogenides (TMDs)  \cite{ZhangHNatureChem, YaoWRev, WangFpseudoBWSe2, XuXD} are good candidates for realizing the R-2 effect due to the large SOC and large site dipole field.  Although hidden spin polarization has been recently reported in the bulk crystal of WSe$_2$ \cite{ KingNatPhys} and predicted in other layered materials, e.g. LaOBiS$_2$ \cite{LaOBiS2NL, ZungerPRB} and (LaO)$_2$(SbSe$_2$)$_2$\cite{LaOSbSe2} films, so far stable semiconducting thin films with large spin polarization induced by the R-2 effect still remain to be realized  experimentally.

\begin{figure*}
\includegraphics[width=16.7 cm] {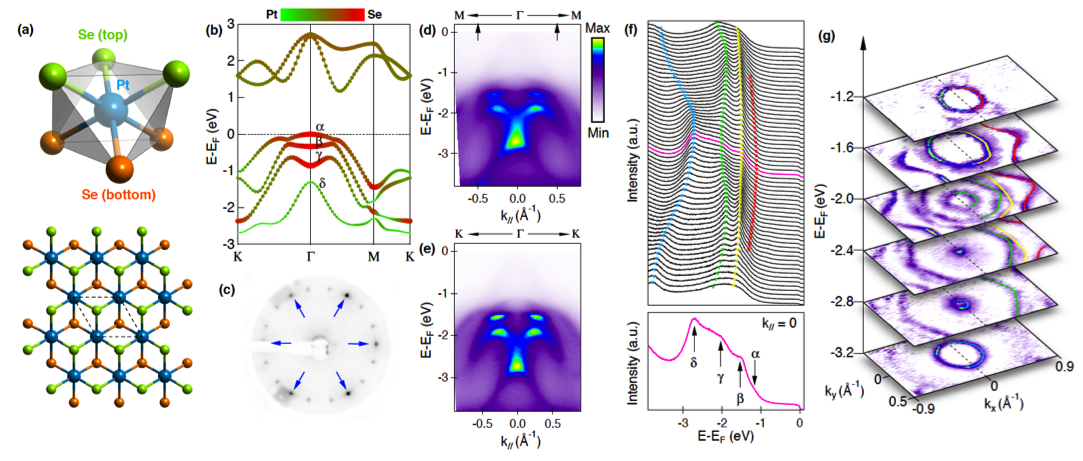}
\caption{\textbf{Crystal structure and electronic structure of monolayer PtSe$_2$.} \textbf{a}, Crystal structure of monolayer PtSe$_2$ (top: unit cell, bottom: top view). \textbf{b}, Band dispersions along the K-$\Gamma$-M-K direction from first-principles calculations. The color and line width distinguish the contribution from Pt and Se (Red for Se and green for Pt).  \textbf{c}, LEED pattern of as-grown monolayer PtSe$_2$ on Pt(111). The peaks from PtSe$_2$ are pointed by blue arrows. Additional diffraction spots rotated from the PtSe$_2$ spots are much weaker and they do not have significant contribute to ARPES measurements. \textbf{d}, ARPES data measured along M-$\Gamma$-M direction. \textbf{e}, ARPES data measured along the K-$\Gamma$-K direction. \textbf{f}, Energy distribution curves (EDCs) for data shown in \textbf{d} for momentum range between the arrows. The EDC at $\Gamma$ point is enlarged at the bottom  panel, from which we can resolve the four bands. \textbf{g}, Constant energy maps at selected energies. The contours of the top four bands at positive k$_x$ are highlighted by colored lines.}
\end{figure*}

Monolayer PtSe$_2$ contains one Pt layer sandwiched between two Se layers, forming trigonal structure  when projected onto the (001) plane (Fig.~2(a)). It has centrosymmetric space group P$\bar{3}$m1 for bulk structure, polar point group C$_3$ for both Pt and Se sites, and it is semiconducting \cite{PtSe2NL}. These properties make it a promising candidate for realizing electrically tunable spintronics by R-2 effect \cite{ZungerNatPhys}. First-principles calculations (Fig.~2(b)) show that the top three valence bands (labeled by $\alpha$, $\beta$ and $\gamma$) are mostly contributed by the \textit{p} orbitals of Se, and the fourth valence band (labeled by $\delta$) is mainly contributed by the \textit{d} orbitals of Pt.  From the calculations, all the bands should be doubly degenerate without any net spin polarization since monolayer PtSe$_2$ has both inversion and time-reversal symmetries. Here, by combining a full three-dimensional spin analysis using spin-ARPES and theoretical calculations, we report the hidden helical spin texture and spin-layer locking in monolayer PtSe$_2$ induced by R-2 effect.

Figure 2(c) shows the low energy electron diffraction (LEED) pattern of the high quality monolayer PtSe$_2$ thin film grown on Pt(111) substrate \cite{PtSe2NL}. The semiconducting property of monolayer PtSe$_2$ has been reported \cite{PtSe2NL}. ARPES data measured along two high symmetry directions M-$\Gamma$-M (Fig.~2(d)) and K-$\Gamma$-K (Fig.~2(e)) show similar dispersions, suggesting that the electronic structure is overall rather isotropic.  Correspondingly, circular shape is observed in the constant energy maps (Fig.~2(g)) for all the bands, and hexagonal warping is observed at relatively high binding energy. Analysis from energy distribution curves (EDCs) shows that the measured dispersions are in good agreement with first-principles calculations (Fig.~2(b)) and the four valence bands are separated from each other at the $\Gamma$ point. 

\begin{figure*}	
\includegraphics[width=12.5 cm] {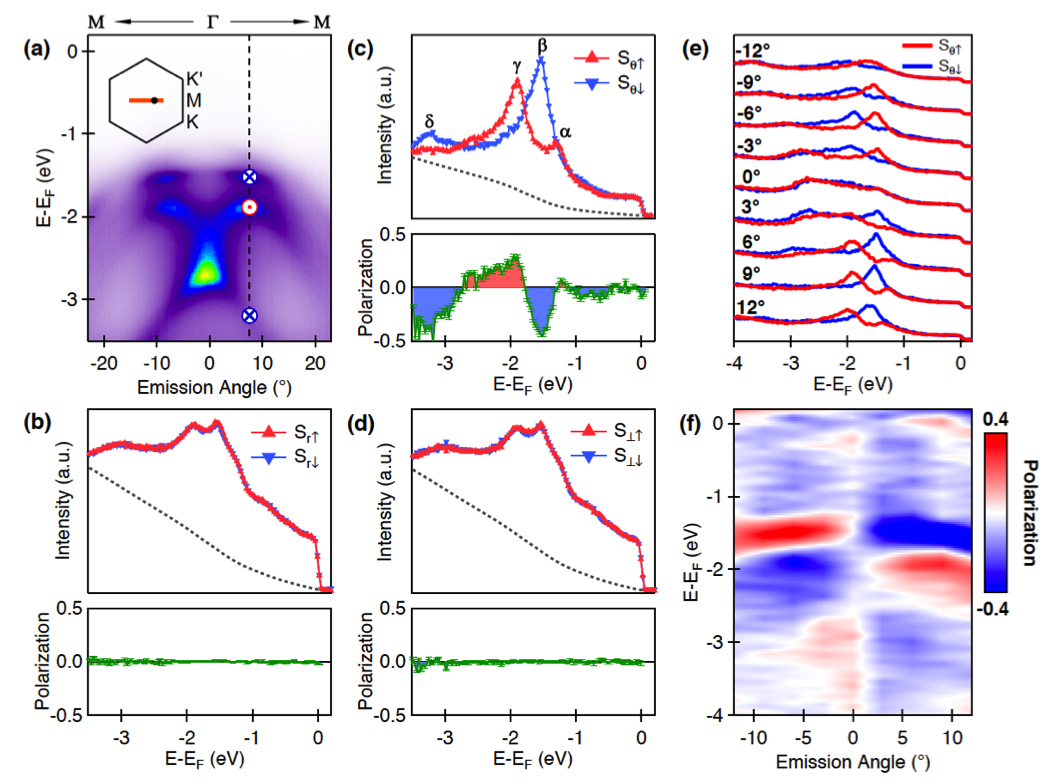}
\caption{\textbf{Spin polarization for the bands along the M-$\Gamma$-M direction.} \textbf{a}, ARPES data measured at different emission angles. The inset shows the position of this cut in the Brillouin zone, and the dot marks the position for EDCs shown in \textbf{b}-\textbf{d}. The red dots and blue crosses stand for the spin directions. \textbf{b}-\textbf{d}, Spin-resolved EDCs of 3 spin components (in-plane radial, tangential directions, and out-of-plane direction) at emission angle of 7.5\textdegree~(along the dashed line in \textbf{a}). The EDCs have been corrected by Shirley background (broken lines). Lower panels are extracted spin polarizations. \textbf{e}, EDCs of spin-up and spin-down states at different emission angles. \textbf{f}, Spin polarization at different emission angles. Red indicates spin-up states, and blue indicates spin-down states along tangential direction.}
\end{figure*}

Figure 3 shows the three dimensional spin analysis for data measured along the M-$\Gamma$-M direction.  A large spin contrast is observed along the tangential direction ($\theta$) for $\beta$, $\gamma$ and $\delta$ bands at emission angle of 7.5\textdegree~(dashed line in Fig.~3(a)) with up to 50$\%$ polarization (Fig.~3(c)), while negligible spin contrast is observed along the radial (r) and out-of-plane ($\bot$) directions (Figs.~3(b,d)). The spin directions are illustrated by blue crosses (into the plane) and red dots (out of the plane) in  Fig.~3(a). We extend the in-plane tangential spin analysis to other momenta along the M-$\Gamma$-M direction.  Figure 3(e) shows the intensity distribution at different emission angles for spin-up and spin-down states along the tangential direction, and the spin polarization is shown in Fig.~3(f). When approaching the $\Gamma$ point from M point, the degree of spin polarization decreases and it becomes negligible at the $\Gamma$ point. On the other side of the $\Gamma$ point, the spin polarization is reversed. The observation of spin polarization along the in-plane tangential direction with opposite signs on the two sides of the $\Gamma$ point suggests that these bands may exhibit helical spin texture.

\begin{figure*}
\includegraphics[width=15.5 cm] {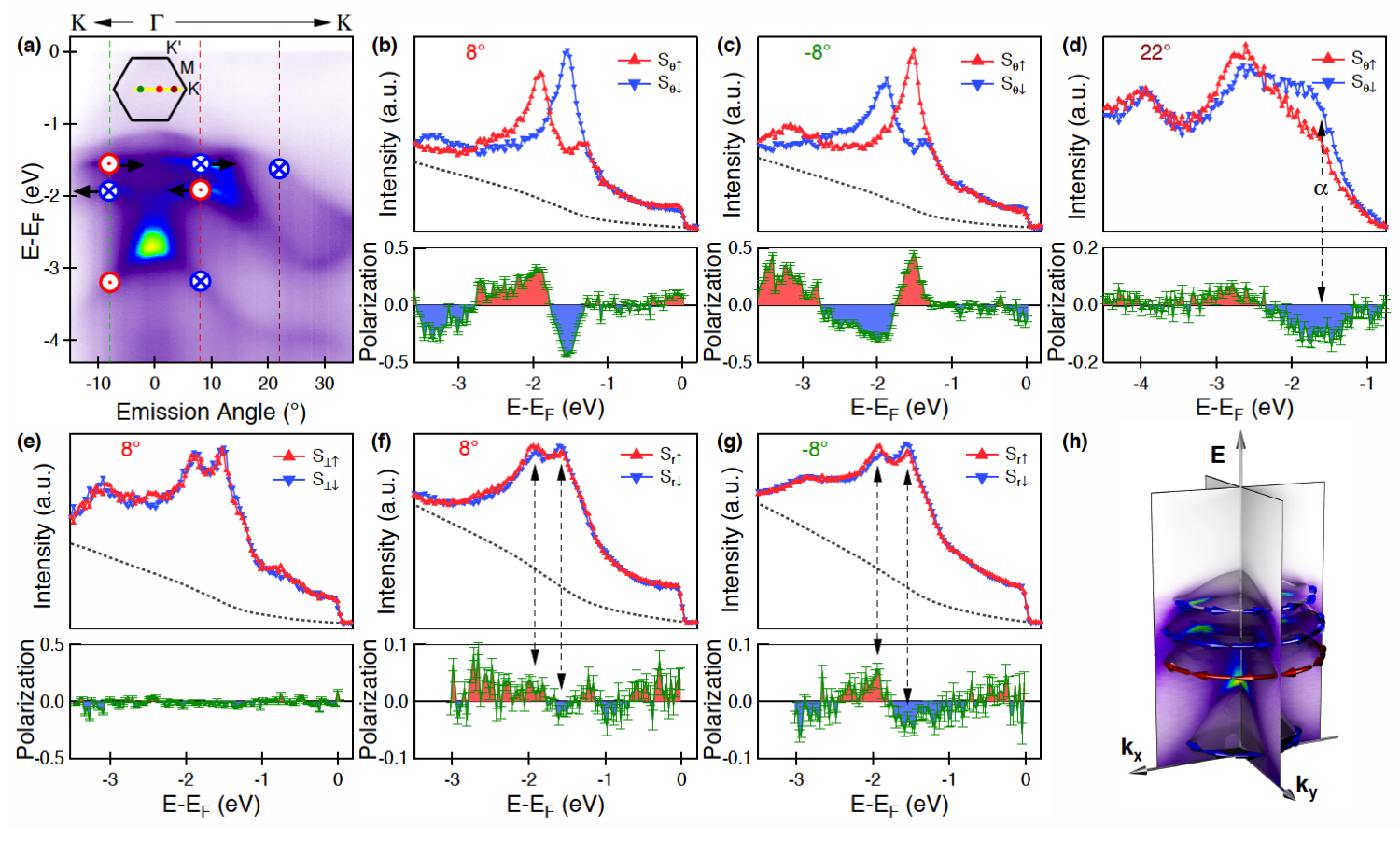}
\caption{\textbf{Spin texture measured along the K-$\Gamma$-K direction and the overall spin texture.} \textbf{a}, ARPES data measured at different emission angles. \textbf{b} and \textbf{c}, Spin-resolved EDCs for the in-plane tangential direction at emission angles of 8\textdegree~and -8\textdegree~ respectively. \textbf{d}, Spin-resolved EDCs for the in-plane tangential direction at emission angle of 22\textdegree. \textbf{e}, Spin-resolved EDCs along the out-of-plane direction at emission angle of 8\textdegree. \textbf{f} and \textbf{g}, Spin-resolved EDCs for the radial direction at emission angles of 8\textdegree~and -8\textdegree~respectively. \textbf{h}, Overview of the spin texture of monolayer PtSe$_2$ from Spin-ARPES measurements.}
\end{figure*}

Figure 4 shows the spin analysis for data measured along the K-$\Gamma$-K direction. A large spin polarization is also observed for the tangential component (Figs.~4(b, c)), in agreement with helical spin texture. The $\alpha$ band shows negligible spin polarization near the $\Gamma$ point. When moving to a large emission angle along the $\Gamma$-K direction, the $\alpha$ band is separated from other bands and we can resolve its spin polarization more easily here. In Fig.~4(d), spin polarization is observed in the $\alpha$ band and its direction is determined to be the same as $\beta$ band. Considering that the electronic structure is rather isotropic, the observed large spin polarization for the tangential direction along both $\Gamma$-K and $\Gamma$-M directions suggests that the spin polarization has an overall helical texture. Different from the $\Gamma$-M direction, a small yet detectable spin polarization of $\approx$ 5$\%$ is observed in the radial component for $\beta$ and $\gamma$ bands with opposite spin directions (Figs.~4(f, g)), and its polarization does not reverse for opposite sides of the $\Gamma$ point. This suggests that the spin structures near the K and K$^\prime$ points in Brillouin zone are not equivalent. This could be related to the defects in sample or other underlying mechanism such as final-state effect. Combining all spin-ARPES measurements discussed above, monolayer PtSe$_2$ on Pt(111) shows overall helical spin textures with different helicities in the four bands as summarized in the schematic drawing in Fig.~4(h).

\begin{figure*}
\includegraphics[width=16.5 cm] {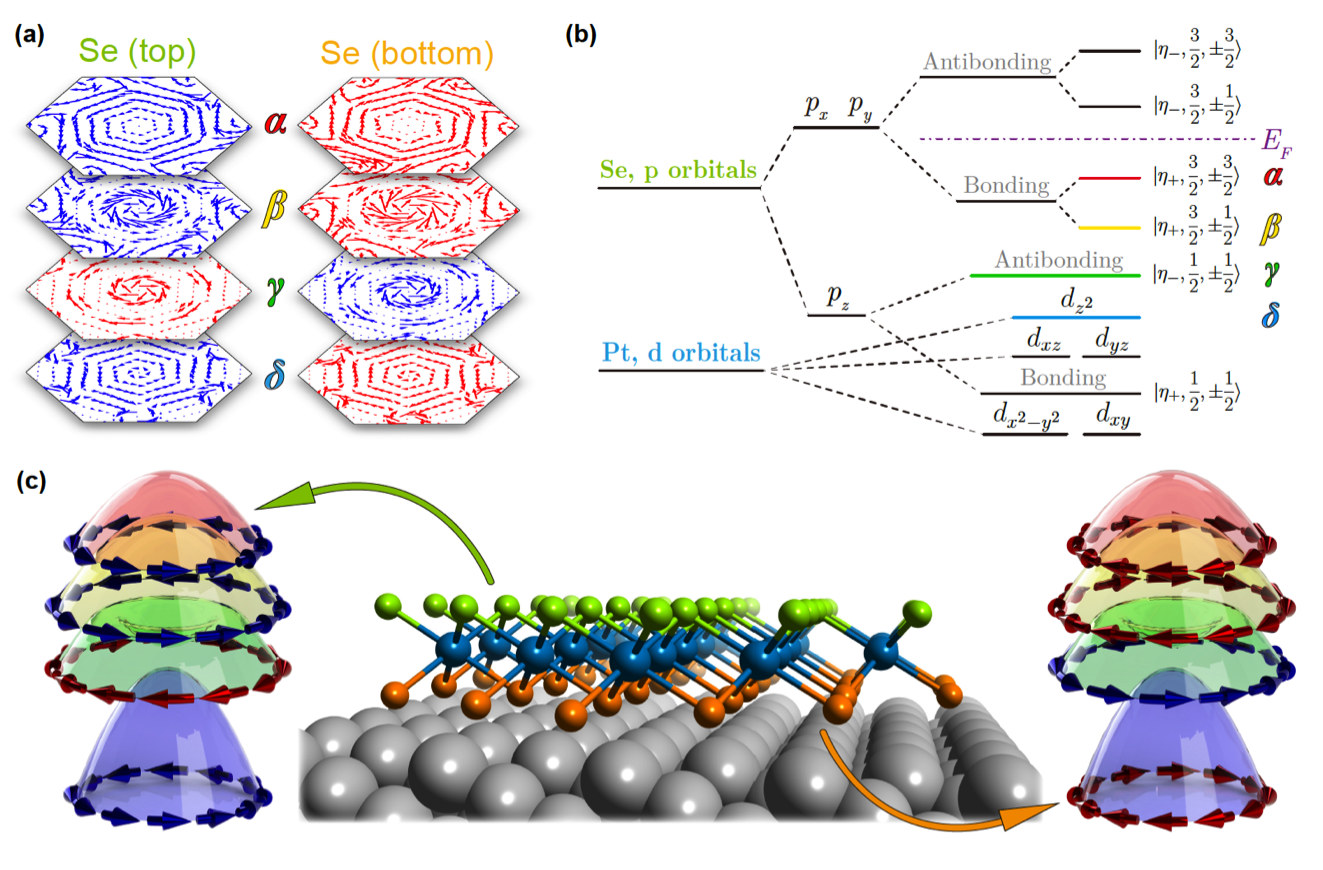}
\caption{\textbf{The unconventional spin-layer locking by R-2 effect proposed for monolayer PtSe$_2$.} \textbf{a}, Spin texture of the two Se layers from first-principles calculation for the four bands respectively. Spins with opposite helicities are spatially separated in the top and bottom Se layers. \textbf{b}, The diagram of Se $p$ orbitals and Pt $d$ orbitals, which dominate the spin texture of bands near Fermi level. \textbf{c}, The schematic diagram for the new spin-locking mechanism of monolayer PtSe$_2$ thin film on Pt(111).}
\end{figure*}

The observed helical spin texture is fundamentally different from that of the R-1 effect.  For the conventional R-1 Rashba effect, two spin-splitting bands possess opposite spin helicities and their eigen-energies become degenerate at the $\Gamma$ point ($k=0$) due to TR symmetry, as shown in Fig.~1(a). In contrast, all these four bands are well split at the $\Gamma$ point, as clearly shown in Fig.~2(f). In additional, the $\alpha$ and $\beta$ bands share the same spin helicities, completely different from the R-1 effect. Compared to the first-principles calculations in Fig.~2(b), it seems confusing that although energy dispersion agrees well between theory and experiments, spin texture does not appear in the calculations due to double degeneracy of two spin bands. This discrepancy can be understood by recognizing the fact that the light for ARPES measurements has a limited penetration depth and thus the top Se layer contributes more significantly to the observed spin polarization. To test this idea, we project spin polarization into the top and bottom Se layers and helical spin textures indeed emerge in each Se layer, as shown in Fig.~5(a). Strikingly, the obtained spin textures on the top Se layer in Fig.~5(a) reproduce qualitative features observed in spin-ARPES measurements for all four bands. For example, the $\alpha$ and $\beta$ bands indeed have the same spin polarization while those of $\gamma$ and $\delta$ bands are opposite. For the $\alpha$ band, the spin polarization is negligible compared to that of the $\beta$ band close to the $\Gamma$ point. The consistency between the calculations and experiments suggests that layer-dependent spin texture (Fig.~5(c)), which was also theoretically proposed in other layered materials including LaOBiS$_2$ \cite{LaOBiS2NL, ZungerNatPhys} and (LaO)$_2$(SbSe$_2$)$_2$ \cite{LaOSbSe2} thin films, plays an essential role in understanding our spin-ARPES results. The existence of spin-layer locking can be understood as a natural consequence of the common sandwich type of crystal structures in all these materials, in which local electric fields are expected to point from the two outer Se layers to the central Pt layer.

More theoretical understanding can be obtained by analyzing the orbital natures of these four bands at the $\Gamma$ point (the atomic limit), as shown in Fig.~5(b). It is found that the $p$ orbitals of Se atoms dominates the bands near the Fermi energy and the $d$ orbitals of Pt atoms are fully occupied. The strong anisotropy introduce a strong energy splitting between the $p_{x,y}$ orbitals and the $p_z$ orbital of Se atoms, pushing Se $p_z$ orbitals to lower energy. As a result, the conduction and valence bands around the band gap mainly consist of the $p_{x,y}$ orbitals of Se atoms. The hybridization of the Se $p_{x,y}$ orbitals between the top and bottom Se layers can be mediated by the central Pt layer and leads to a band gap opening between the conduction and valence bands, which are formed by the bonding and anti-bonding states of Se $p_{x,y}$ orbitals, respectively. After taking into account SOC, we find that the $\alpha$ and $\beta$ bands correspond to the bonding states of Se $p_{x,y}$ orbitals with total z-direction angular momentum $\pm\frac{3}{2}$ and $\pm\frac{1}{2}$, respectively. We explicitly show the atomic orbital form of the basis wave function and construct the corresponding low energy effective Hamiltonian for the $\alpha$ and $\beta$ bands based on the symmetry principle in the Supplementary Information. The effective Hamiltonian clearly shows layer dependent spin textures for the $\beta$ bands. In addition, it is found that the vanishing spin texture of the $\alpha$ band at a small momentum results from the $\pm\frac{3}{2}$ angular momentum, in contrast to the $\pm\frac{1}{2}$ angular momentum of the $\beta$ band. 

In summary, for the first time, we report the experimental realization of unconventional R-2 Rashba effect in a centrosymmetric monolayer PtSe$_2$ thin film. Considering that the monolayer PtSe$_2$ sample is very stable, only monolayer thick and semiconducting, growth or transfer of such thin film onto insulating substrates may provide exciting opportunities to realize electrically controllable spintronics devices.

~\\
{\large{\bf Methods}}
\\{\bf Experiments.}
The PtSe$_2$ thin film was grown by direct selenization of Pt(111) \cite{PtSe2NL}. The growth stops when the substrate is covered by one monolayer of PtSe$_2$. Spin-ARPES measurements were performed at ESPRESSO endstation \cite{ESPRESSO} of HiSOR. The normal and spin ARPES measurements were both taken at 20 K, using photon energies of 21.2 eV (UV lamp) and 20 eV (Synchrotron radiation). These two light sources give the same spin structures. Spin polarizations are calculated by $P=A/S_{eff}$, where $A=(I_{+}-I_{-})/(I_{+}+I_{-})$ is the intensity asymmetry for different magnetization directions of the detector target, and $S_{eff}$ is the effective Sherman function for the spin detector. For radial and out-of-plane components, the effective Sherman function is 0.3, and for tangential component, the effective Sherman function is 0.235.
\\{\bf Calculations.}
The first-principles calculations are performed using the density functional theory as implemented in the Vienna ab initio simulation package \cite{VASP} with the projector augmented-wave method. Perdew-Burke-Ernzerhof parametrization of the generalized gradient approximation is used for the exchange-correlation potential \cite{PBE}. We adopt a default plane-wave energy cutoff, and the Brillouin zone is sampled by a $\Gamma$-centered $6\times6\times1$ $k$-point mesh. The monolayer structure of PtSe$_2$ is modeled with a vacuum region more than 15 \AA~thick to eliminate the spurious interaction between neighboring layers. A perpendicular electric field of 0.1 eV/\AA~was applied to simulate the effect of the substrate. Spin-orbit coupling (SOC) is included in all electronic structure calculations.

\bibliography{RefSpinPtSe2}
\bibliographystyle{naturemaga}
~\\
{\large{\bf Acknowledgments}}
\\This work is supported by the National Natural Science Foundation of China (Grant No.~11274191, 11334006 and 11427903), Ministry of Science and Technology of China (Grant No.~2015CB921001).  Spin-ARPES experiments at HiSOR have been performed under the proposal No. 14-A-15.

~\\
{\large{\bf Author Contributions}}
\\S.Z. designed the research project. W.Y. and K.D. prepared the samples. W.Y., E.W.,  K.D., M.Y., K.Z. and S.Z. performed the spin-ARPES measurements and data analysis with assistance from T.O.  H.H. and W.D. performed the first-principles calculations and C.L. worked out the effective analytical model. W.Y., C.L. and S.Z. wrote the manuscript, and all authors commented on the manuscript.

\end{document}